%% file: doleqm.tex
\documentclass[11pt, letterpaper, twocolumn]{article}

\usepackage{amssymb}
\usepackage{amsmath}  
\usepackage{amsfonts} 
\usepackage{graphicx} 
\usepackage{abstract}
\usepackage{xfrac}
\usepackage{parskip}

\usepackage[letterpaper, total={7.2in, 9.5in}]{geometry}

\usepackage[font={small,it}]{caption}

\usepackage{import}
\usepackage{xifthen}
\usepackage{pdfpages}
\usepackage{transparent}

\let\sm\scriptstyle

\begin{document}

\title{
  \bf{
    {\huge{Snakes on a Hyper-Plane:\\}}
    {\large{Dynamical and Ontological Locality in Everettian Quantum Mechanics}}}}
\author{Travis Norsen \\ Smith College \\ tnorsen@smith.edu}
\date{\today}

\twocolumn[
\maketitle 
\begin{onecolabstract}
  \noindent \emph{An apparently emerging consensus holds that Everettian Quantum
    Mechanics (EQM), in the Spacetime State Realism (SSR) version of Wallace
    and Timpson, involves ontological non-locality (i.e.,
    non-separability) but is fully compatible with the dynamical
    notion of locality usually taken to be required by relativity.
    The claim of dynamical locality is based on the fact that the
    local beables postulated by the theory (namely, Reduced Density
    Matrices, RDMs,
    associated with localized regions) are not affected by
    interventions outside their past light cones.  However, the theory's
    non-separability means that it posits non-local
    beables (namely RDMs for non-localized regions) in addition to the local
    ones.  We argue here that the involvement of non-local beables
    introduces complications for extant formulations of dynamical
    locality and at least suggests, contrary to the emerging consensus,
    that SSR EQM should perhaps not be considered compatible with
    relativistic causality after all.  }  \\
\hspace{.2in}\\
\end{onecolabstract}
]

\maketitle

\section{Introduction}

Back in the 1990s, the distinction between \emph{dynamical
  non-locality} (i.e., “action-at-a-distance” or more precisely
“faster-than-light causal influence”) and \emph{ontological
  non-locality} (i.e., “non-separability” or “irreducible holism”)
received a lot of attention in the context of Bell’s theorem.  This
was motivated by Jarrett’s suggestion \cite{jarrett} that Bell’s notion of local
causality could be decomposed into two sub-conditions that Jarrett
originally dubbed ``locality'' and ``completeness'', but which eventually
came to be more widely known by Shimony’s more neutral terminology of
``parameter independence''
(PI) and ``outcome independence'' (OI).
This PI/OI distinction was then (somewhat implausibly) assimilated onto the
dynamical/ontological locality distinction
such that violations of PI were regarded as incompatible with
relativistic causality,
whereas violations of OI were regarded as merely indicating a novel
sort of ``non-separability''
or ``irreducible holism'' that could enjoy peaceful coexistence
with relativity. \cite{graybook}

\begin{figure}[t!]
  \centering
    \def\svgwidth{\columnwidth}
    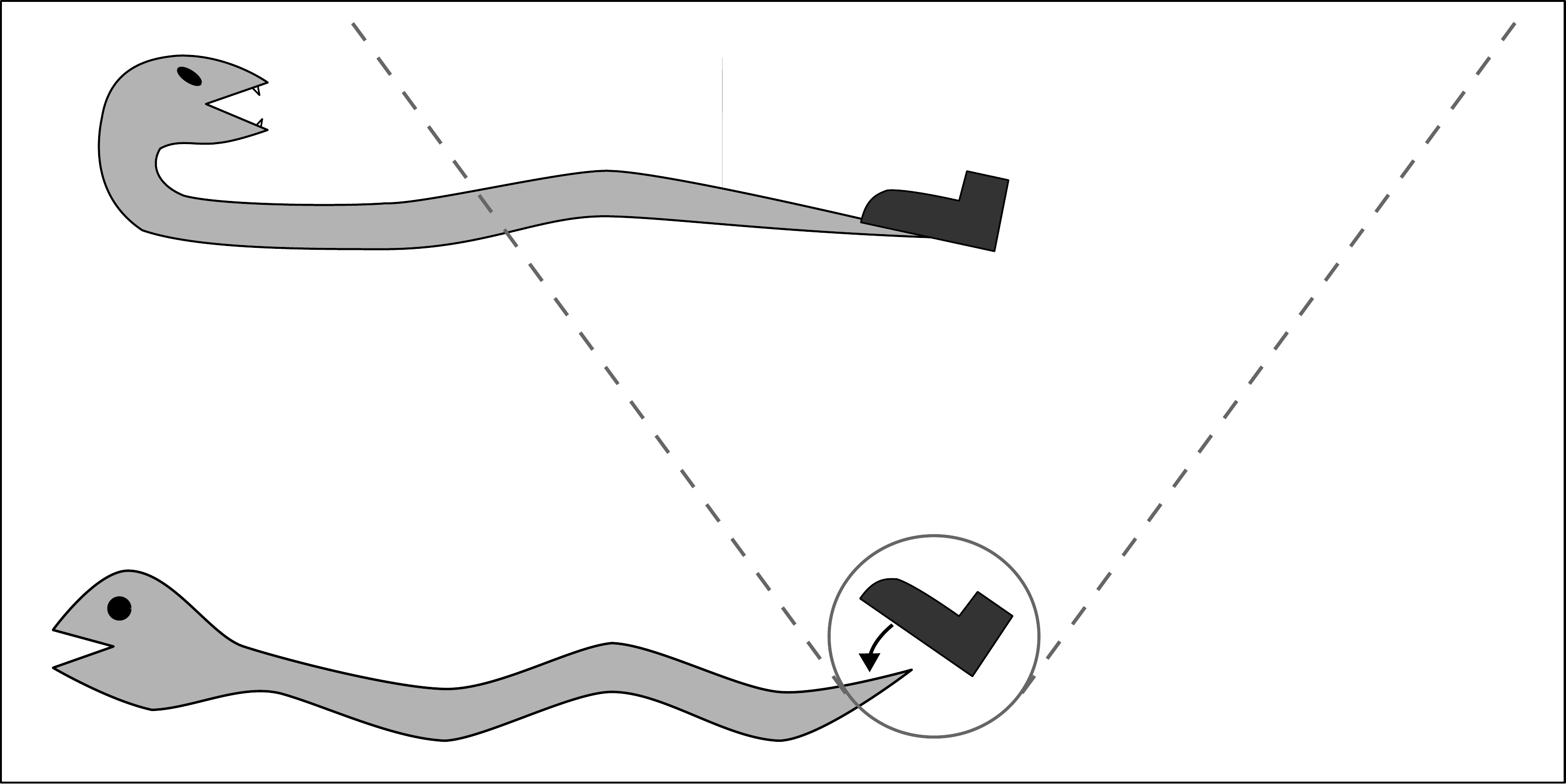

  \caption{Pseudo-space-time diagram of Arthur Fine's dramatic example:  If you step on a snake's
    tail and its head rears around in response, outside the future
    light cone of the stepping (i.e., sooner than would be allowed if
    the information about the stepping propagated to the head at the
    speed of light), there is a violation of relativistic dynamical
    locality whether or not the snake should be considered, in some
    sense, a non-separable, irreducible whole.
        }
  \label{fig-snake1}
\end{figure}

While I was a graduate student in physics at the University of
Washington (between 1997 and 2002), just getting interested in these
sorts of issues, I was lucky to have the opportunity to learn from and
discuss with Arthur Fine, who moved to UW's philosophy department
(from Northwestern) during this period.  We didn’t see eye to eye on
everything, but we did strongly agree that this idea -- of supposedly
avoiding a conflict with relativistic dynamics by citing non-separable ontology --
seemed suspicious.  And I have always remembered the dramatic way
Prof. Fine expressed his skepticism about this once in conversation:
``If you step on a snake’s tail, and its head immediately rears around
to prepare to bite you, does it really matter whether you say a causal
influence propagated along the length of the snake or instead regard
the snake as an irreducible whole?''  (See the cartoon depiction in
Figure \ref{fig-snake1}.)

Fine’s point, of course, was that either way the rearing around of the
snake's head is a real effect, caused by the tail being stepped on.
And if the effect is outside the future light cone of the cause, this
should be regarded as conflicting with relativistic causality
regardless of how we might want to further analyze the metaphysics
involved.
A theory with ontological
non-locality (i.e., non-separability, irreducible holism) can \emph{also} be
dynamically non-local, with the ``irreducible wholes'' perhaps
explaining -- but not explaining away -- the faster-than-light
influences. 

Fine's insight, I think, proved correct and the initial excitement over
achieving ``peaceful coexistence'' between quantum mechanics and
relativity, in this particular way, waned as people began to
appreciate the dubiousness both of decomposing Bell’s local causality
into PI and OI, and of associating violations of OI with
non-separability as opposed to dynamical non-locality.  \cite{bvj}

However, similar themes have -- so to speak -- reared their heads
again more recently in discussions of the status, vis a vis both
dynamical and ontological locality, of Everettian Quantum Mechanics (EQM).
Pointed questions about locality have arisen in particular about the
process of branching -- into decohered worlds -- that occurs, for
example, when measurements are performed.  In the standard EPR
setup, for instance, spatially-separated experimenters Alice and Bob
share a pair of (say) spin-$\sfrac{1}{2}$ particles in the singlet state; if Alice
now measures (say) the z-spin of her particle, the overall quantum
state evolves into a superposition of terms which Everettians
interpret as a description of two overlapping but (due to decoherence)
subsequently non-interacting worlds, one containing an Alice who saw
``spin up'' and one containing an Alice who saw ``spin down''.

But does Alice's measurement cause a splitting not only of Alice and
her particle, but also of distant Bob and/or his particle?  Those who
say ``yes'' and thereby endorse a ``global branching'' view have tended to
see the theory as violating relativistic dynamical locality.  But
it is also possible to say ``no'' and thus adopt a ``local branching'' view
that appears more compatible with dynamical locality.   (See
\cite{ney} for a clear discussion of the lay of the land here.) 

Pressing further, though, we might consider the case where both Alice
and Bob perform z-spin measurements of their particles
(simultaneously, let’s say, in some Lorentz frame).  Local branchers
will say that both Alice and Bob split.  And full locality would
evidently require these two (localized, space-like separated)
splitting events to be fully independent.  But they are not:  the
world containing the Alice who sees ``spin up'' inevitably also contains
the Bob who sees ``spin down'', and likewise ``spin down Alice'' and ``spin
up Bob'' also inhabit the same world.  So there is some appearance of
non-locality in the \emph{correlatedness} of distant localized
branchings.  (The worry that this correlatedness implies a violation
of relativistic causality has been expressed, for example, in
\cite{solipsism}, \cite{tumulka}, and \cite{cw}.)

A seemingly emerging consensus view of this situation concedes the
appearance of non-locality but claims that it is purely of the
(dynamically benign) ontological variety.  That is, the consensus view
sees Everettian quantum mechanics as exhibiting ontological
non-locality -- i.e., non-separability --  but maintaining perfectly local
dynamics.

My goal in this paper is simply to question this developing consensus.
In particular, I will argue that existing formulations of dynamical
locality basically presuppose ontological locality, and that, as soon
as we consider theories with non-separability, some novel
possibilities arise for the space-time relations between causes and
effects which at very least raise difficult questions for, and may
perhaps be seen as undermining, the emerging consensus view. 

Before jumping into a detailed discussion, let me try to put the
central issue on the table using a modification of Fine’s snake
example.   

In the original example, the ``effect'' event (the rearing around of
the snake’s distant head) is sufficiently localized that it is fully
outside of the future light cone of the also-localized ``cause'' event
(the nearby stepping on the snake’s tail).  This is why, regardless of
any further details one might want to fill in regarding the
precise metaphysics and causality involved in the connection between
the two events, we have an unambiguous violation of dynamical locality.

\begin{figure}[t!]
  \centering
    \def\svgwidth{\columnwidth}
    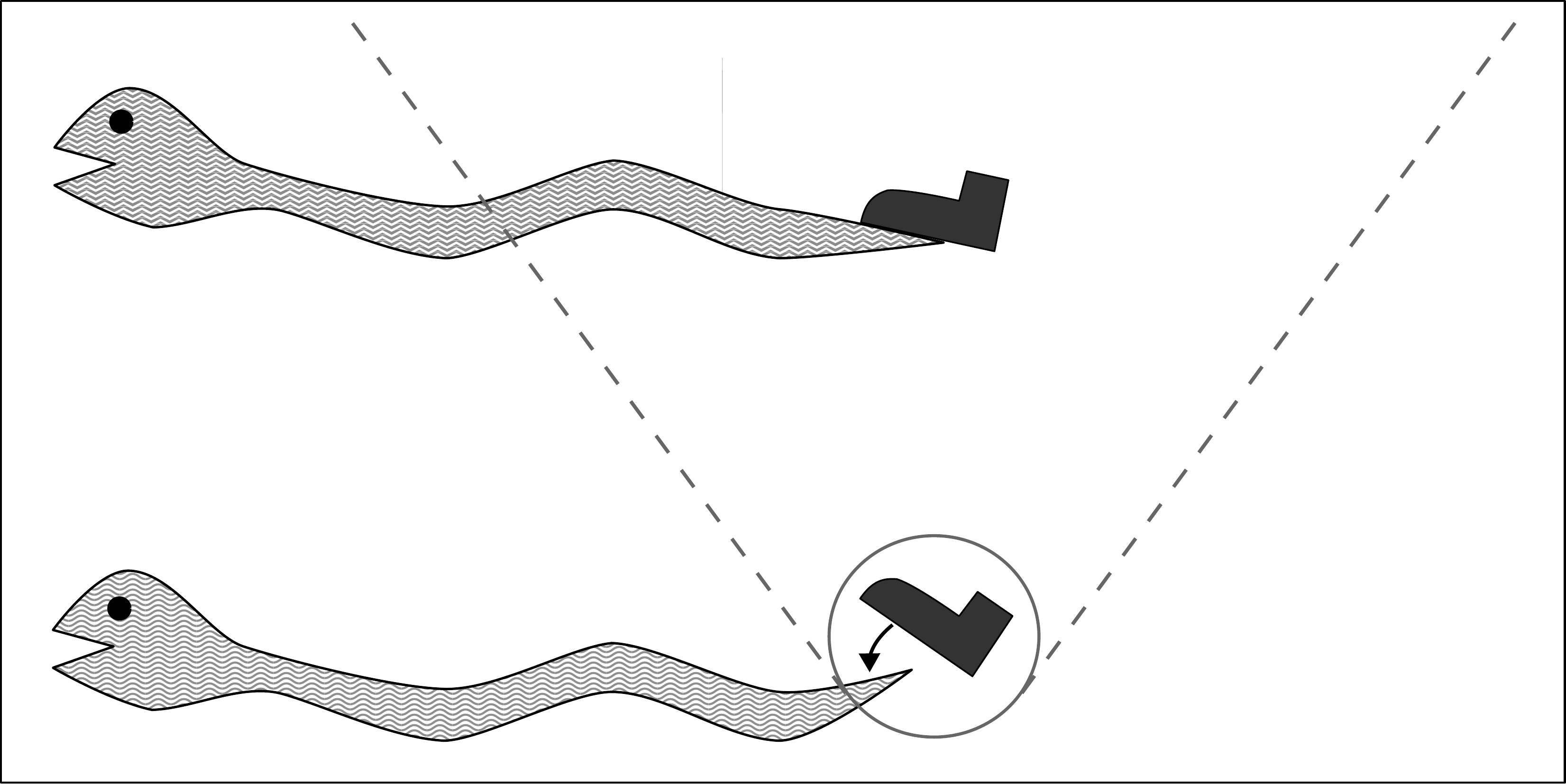

  \caption{Suppose, modifying Fine's example, that stepping on the snake's tail
    does not affect any properties of the snake that are localized in
    its head (or any other part), but instead only changes a global, irreducibly holistic
    property (misleading depicted here as a change from a wavy to a
    more spiky stripe pattern).  Should this be considered compatible
    with dynamical locality?
        }
  \label{fig-snake2}
\end{figure}

But by definition in a theory with some non-separability we have
objects/events which are irreducibly de-localized.  Let's consider the
implications of this by pretending that, in
addition to localized properties such as the shape of and pressure
exerted on their tails and the orientation of their heads, snakes
also possess a mysterious holistic property which can be
affected by an external intervention.  In Figure
\ref{fig-snake2} I have tried to capture the idea visually in terms of
the snake's global pattern of stripes changing from wavy to spiky when
its tail is stepped on.  This visualization,
of course, is misleading.  An actual wavy stripe pattern, for example,
is entirely reducible to facts about the snake's skin coloration on
each of its constituent parts.  Whereas in the example we mean to
consider here, the property in question is irreducibly global,
holistic.   It exists, somehow, throughout the entire snake-shaped
region occupied by the snake, but is not contained in any part of that
region and is not
reducible to or grounded in \cite{ground} even a
complete accounting of the snake's localized properties.  So the
example supposes that when we step on the snake's tail, no localized
properties of the snake (outside the tail region) change.  Only
this non-localizable, irreducible, holistic property
changes.

So, in our pretend example, a localized ``cause'' event
(stepping on the snake’s tail) triggers an ``effect'' event (the change
of the snake’s holistic global property from wavy to spiky)
which
exists, irreducibly, across the entire spatial extent of the snake.
But this effect, as shown in Figure \ref{fig-snake2}, is thus both
partially inside of
and partially outside of the future light cone of the cause.

Should
this be considered compatible with dynamical locality (perhaps on the
grounds that the effect is at least partially in the future light
cone of the cause, or perhaps on the grounds that the cause is fully within the
past light cone of the effect)?  Or should it instead be considered
incompatible with dynamical locality (on the grounds that the effect
is not \emph{confined} within the future light cone of the cause)?

That is the key question we will develop in the rest of the paper.
We begin in the following section by reviewing several relevant
formulations of dynamical locality and exploring in more detail the novel
possibilities that arise for theories that posit (certain sorts of
``snake''-shaped) non-local beables in
addition to local beables.  Section 3 lays out the ontology postulated
in the Spacetime State Realism (SSR) version of Everettian Quantum
Mechanics (EQM)
and shows how it exhibits (at least) one of the novel behaviors that I
think might plausibly be considered a case of dynamical nonlocality.  Section 4
summarizes the result and connects it with recent discussions of
branching and relational properties.

\section{Dynamical Locality and local/nonlocal beables}
\label{sec2}

In his classic 1990 paper \emph{La Nouvelle Cuisine}, Bell begins his
discussion of dynamical locality by
noting that, in order to judge the compatibility of a candidate theory
with the relativistic idea that causal influences cannot go faster
than light, one must first be clear about the ontology, in ordinary 3D
space / 4D space-time, postulated by the theory:
\begin{quote}
``...you must identify in your theory `local \emph{be}ables'.  The
\emph{be}ables of the theory are those entities in it which are, at
least tentatively, to be taken seriously, as corresponding to
something real. ....  \emph{Local} beables are those which are
definitely associated with particular space-time regions.''  \cite{bell90}
\end{quote}
Familiar examples from classical theories include the particles
(with definite locations in space at each moment in time) of classical
mechanics or the fields (with definite values at each point in
space-time) of electromagnetism.

Note that \emph{all} of the beables
postulated by pre-quantum theories are \emph{local} beables, i.e.,
beables associated with localized (perhaps pointlike) regions.
Despite being in some sense delocalized, facts
pertaining to larger regions (such as the total energy in the
electromagnetic field in some extended region) should not be thought
of as non-local beables, but rather simply as emergent
(non-fundamental) properties that are fully reducible to local beables
and their properties.\footnote{Here it's possible I disagree slightly
  with Bell, who says that ``the total energy in all space'', for
  example, ``may be a beable, but is certainly not a local one.''  But
my guess here is that Bell was speaking loosely, trying to provide an
example of a non-local beable from a familiar, classical theory when,
in fact, there are none. \cite{bell90}}

\begin{figure}[t!]
  \centering
    \def\svgwidth{\columnwidth}
    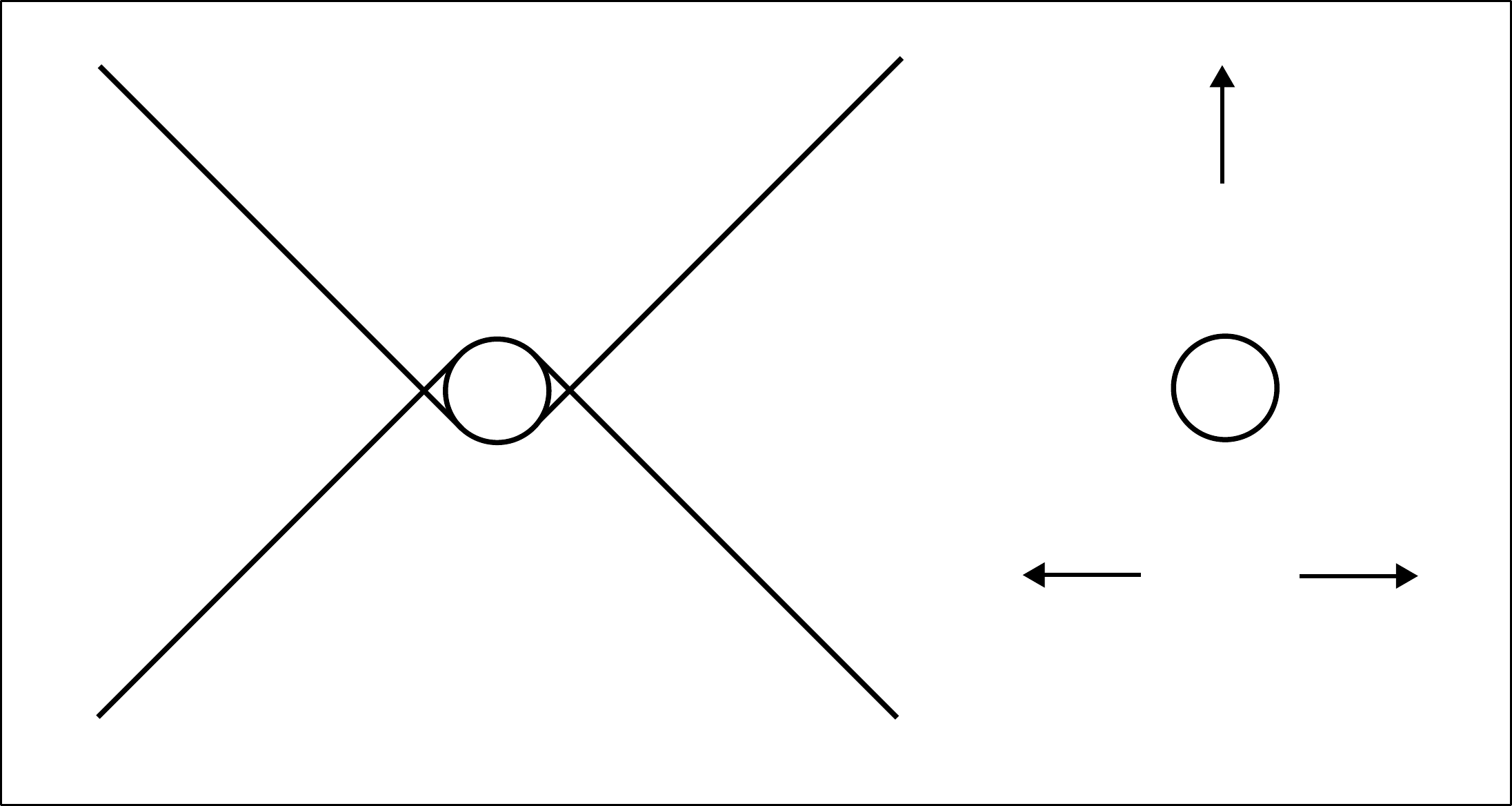

  \caption{``Space-time location of causes and effects of events in
    region 1.''  \cite{bell90}
        }
  \label{fig-plc}
\end{figure}

\begin{figure}[t!]
  \centering
    \def\svgwidth{\columnwidth}
    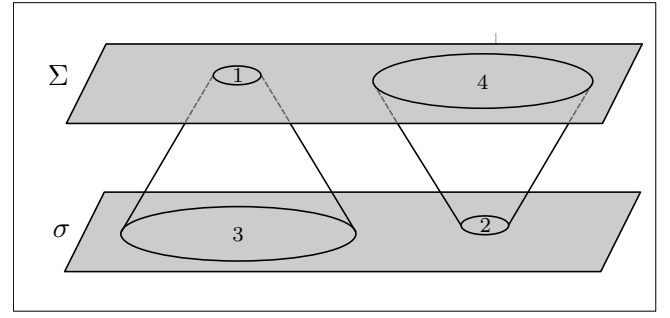

  \caption{Diagram for the several formulations of dynamical
    locality.  $\sigma$ and $\Sigma$ are, respectively, earlier and
    later space-like hyperplanes.
    $3$ is the intersection of the
    past light cone of $1$ with $\sigma$, while $4$ is the
    intersection of the future light cone of $2$ with $\Sigma$.  $3$
    and $2$ are non-overlapping, so $2$ is at space-like separation
    from $1$.  According to Bell's formulation of dynamical locality:
    ``Full specification of what happens in 3 makes events in 2
    irrelevant for predictions about 1 in a locally causal theory.''  \cite{bell90}
        }
  \label{fig-lc}
\end{figure}

Bell explains the principle of relativistic causality as follows,
referencing the diagram we have reproduced in Figure \ref{fig-plc}:
\begin{quote}
``...for events in a space-time region 1 ... we would look for causes in
the backward light cone, and for effects in the future light cone.  In
a region like 2, space-like separated from 1, we would seek neither
causes nor effects of events in 1.''  \cite{bell90}
\end{quote}
He then suggests the following more ``sharp and clean'' formulation of
this idea, referencing a simpler version of our Figure \ref{fig-lc}:
\begin{quote}
``A theory will be said to be locally causal if the probabilities
attached to values of local beables in a space-time region 1 are
unaltered by specification of values in local beables in a space-like
separated region 2, when what happens in the backward light cone  of 1
is already sufficiently specified, for example by a full specification
of local beables in a space-time region 3....'' \cite{bell90}
\end{quote}
Bell speaks of probabilities to ensure that the formulation captures
the relativistic idea of ``no faster-than-light causal influences''
even for irreducibly stochastic (i.e., non-deterministic) theories.  In this case, we might
express Bell's local causality condition mathematically as follows:
\begin{equation}
  P(b_1 | B_3, b_2) = P(b_1 | B_3)
  \label{eq-bellloc}
\end{equation}
where we use a lowercase ``$b"$ to indicate the value of some
particular local beable in the region in question and a capital
``$B$'' to denote a complete specification of \emph{all} beables in the
region in question.

Note that while region $2$ is outside the past
light cone of region $1$, part of $4$ may be inside the future light
cone of $3$.  So even though $4$ is at space-like separation from $1$,
Equation \eqref{eq-bellloc} -- but with $2$ replaced by $4$ -- would
no longer be reasonable to demand of a locally causal stochastic
theory:  $4$ could contain information about random events subsequent
to $\sigma$ and not determined by $B_3$ which locally influence
both $1$ and $4$.   (See \cite{bellsconcept} for more discussion.)  So
some care is required to formulate dynamical locality appropriately
for truly general (not necessarily deterministic) theories. 

But we will be primarily
concerned with Everettian Quantum Mechanics (EQM), which is 
deterministic.  And for deterministic theories, it is much simpler to
capture what dynamical locality requires:  $b_1$ should be a function
of $B_3$, with anything else from $\sigma$ (such as some $b_2$) being
redundant/irrelevant:  $b_1(B_3,b_2) = b_1(B_3)$.  Or we could
equivalently express the idea this way:
\begin{equation}
b_1(B_3, b_2) = b_1(B_3,b'_2)
\end{equation}
where $b_2$ and $b'_2$ are two different possible values of some
beable in region 2.

Note that this requirement captures both aspects of the qualitiative
version of locality depicted in the earlier Figure:  the \emph{causes} of
events in (Figure \ref{fig-lc}'s) region 1 are confined to its past light cone, which
excludes events in  region $2$; and the
\emph{effects} of events in $2$ are confined to its future light cone,
which excludes events in $1$.

These two aspects of dynamical locality have
been helpfully formulated in a recent paper by Lev Vaidman.
\cite{lev}

Vaidman's ``forward'' version of dynamical locality (let's call it $F$)
captures, at least for deterministic theories, the idea that
the effects of a given event should be confined to that event's future
light cone.  Taking $2$ (from Figure \ref{fig-lc}) as an ``action region'' (which we can think of
as a place where some possible ``intervention'' might occur) dynamical locality
demands that two possible evolutions which differ, on $\sigma$, only
within region 2 (i.e., they are identical in $\bar{2}$, the complement of
$2$ on $\sigma$) can differ, on $\Sigma$, only within region $4$.
Thus $b_4(B_{\bar{2}},B_2)$ need not necessarily equal
$b_4(B_{\bar{2}},B'_2)$ (where $B_2$ and $B'_2$ are two different
complete specifications of beables in region $2$).  But dynamical
locality requires
\begin{equation}
  b_{\bar{4}}(B_{\bar{2}},B_2) = b_{\bar{4}}(B_{\bar{2}},B'_2),
  \label{forward}
\end{equation}
i.e., if two evolutions differ only inside of some region $2$ on $\sigma$, they
must agree outside of the future light cone of $2$, i.e., outside of
region $4$, on $\Sigma$.  

And then Vaidman's ``reverse'' version ($R$) of dynamical locality (also
articulated and applied recently in the paper by Chua and Sebens \cite{cs}) captures
(again, at least for deterministic theories) the idea that the causes
of a given event should be confined to that event's past light cone.
Taking now $1$ as an ``effect region'' the idea is that, considering
again two possible evolutions, differences in
region $1$ of $\Sigma$ can arise only if there are differences within
region $3$ of $\sigma$.  That is, if the two evolutions differ on
$\sigma$ only outside of region $3$ (i.e., only in $\bar{3}$), they
must be identical in $1$.  So dynamical locality requires that
\begin{equation}
  b_1(B_3, B_{\bar{3}}) = b_1(B_3,B'_{\bar{3}}),
  \label{reverse}
\end{equation}
i.e., if two evolutions differ only outside of some region $3$ on $\sigma$, they
must agree within the future light cone of $3$.  

It should appear that Vaidman's ``forward'' and ``reverse'' versions
of dynamical locality -- $F$ and $R$ -- are equivalent to each other.  After all, $1$ is
just some
particular location within $\bar{4}$, which is therefore outside the future
light cone of $2$.  But then $2$ is outside the past light cone of
$1$, i.e., is just a particular location within $\bar{3}$.  So
Equation \eqref{forward} (understood as applying, for an arbitrary
given $2$ on $\sigma$, to all possible points within $\bar{4}$ on $\Sigma$) is indeed
equivalent to Equation \eqref{reverse} (understood as applying, for an
arbitrary given $1$ on $\Sigma$, to all possible points in $\bar{3}$
on $\sigma$).  And both appear equivalent to Bell's formulation as
well (again, at least for deterministic theories). 

The crucial point I want to introduce, though, is that all of these equivalent
formulations of dynamical locality tacitly assume that we are dealing with a theory (like
those familiar from classical physics) in which all beables are local
beables -- a TELB (``theory of exclusively local beables''
\cite{telb}).\footnote{It has been pointed out before
that although Bell's formulation of dynamical locality
basically presupposes a TELB, we could supplement $B_3$ in Equation
\eqref{eq-bellloc} with additional stuff -- which could include
local beables from $\bar{3}$ but could also include non-local beables somehow
associated with $\sigma$ that are neither fully in $3$ nor $\bar{3}$
-- and still regard Equation \eqref{eq-bellloc} as a necessary
condition for (rather than a definition of) dynamical locality.
\cite{bellsconcept,scholarpedia}
But the
events $b_1$ and $b_2$ of Bell's formulation -- and the events
in the 
``action'' and ``effect'' regions in Vaidman's formulations -- have
traditionally been thought of as
aspects/summaries of things like measurement outcomes or settings on a
measuring apparatus -- things, that is, that are (or are reducible to)
local beables.  The possibility of non-local beables being among the
considered causes and effects (rather than the background we use to
screen off non-causal correlations) has not, to my knowledge, received
much attention, no doubt largely because non-local beables are cryptic
and unfamiliar.  But theories which posit non-local beables do so for
a reason, namely, that they play an ineliminable role in the dynamical
(cause-and-effect) relations described by the theory.  So the
non-local beables deserve to be taken seriously and their role in the
dynamics deserves to be scrutinized.}
Basically we have assumed that each beable contributing
to a complete description of the physical state on a given hyperplane
is either unambiguously inside of, or unambiguously outside of, a
given region.

In particular, $F$ presumes that beables on
$\Sigma$ are either unambiguously in $4$ (and hence permitted to vary
in response to variations confined to $2$) or instead in $\bar{4}$ (and
hence not permitted to vary in response to variations confined to $2$).
Similarly, $R$ presumes that beables on $\sigma$ are either
members of $B_3$ (variation of which may produce changes in $1$) or
instead members of $B_{\bar{3}}$ (variation of which may not produce
changes in $1$).

\begin{figure}[t!]
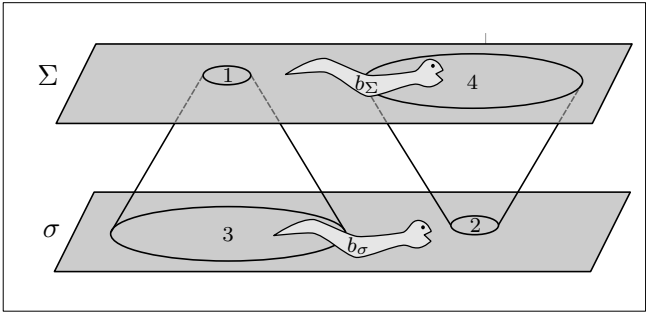

  \centering
    \def\svgwidth{\columnwidth}
    

  \caption{In a theory with snake-shaped non-local beables, our
    earlier (tacitly TELB-based) formulations of locality become
    ambiguous.  For example:  Is a snake-shaped non-local beable such
    as $b_\Sigma$ (that is partly inside of, but also partly outside
    of, the future light cone of region $2$) allowed to be among the
    effects of causes in $2$?  Is $b_\sigma$ (partly inside of, but
    also partly outside of, the past light cone of region $1$) allowed
    to be among the causes of effects in $1$?
        }
  \label{fig-newsnakes}
\end{figure}

But non-local (e.g., snake-shaped) local beables can by definition
span the boundary between $3$ and $\bar{3}$ or between $4$ and
$\bar{4}$, and hence render our previous formulations of dynamical
locality ambiguous.  Two cases of special interest are illustrated in Figure
\ref{fig-newsnakes}:
\begin{itemize}
\item Case (i):  A non-local beable such as $b_\Sigma$, which
  spans the boundary between $4$ and $\bar{4}$, is among the effects of
  a localized intervention in $2$.

\item Case (ii):  A non-local beable such as $b_\sigma$, which spans
  the boundary between 
  $3$ and $\bar{3}$, is among the causes of a
  localized effect in $1$. 
\end{itemize}
Both cases raise the novel question:  is
this compatible with dynamical locality?  There are two
relatively-obvious possible ways of answering.  

A \emph{permissive} view would answer ``yes''.  That
is, it would say that
\begin{equation}
  b_\Sigma (B_2,B_{\bar{2}}, B_\sigma) \ne b_\Sigma(B'_2,B_{\bar{2}},B_\sigma)
  \label{eq-perm1}
\end{equation}
(where $B_\sigma$ is a complete specification of any non-local beables
on $\sigma$ that span the $2-\bar{2}$ boundary) should \emph{not} count as a violation of local causality since $b_\Sigma$ is
at least \emph{partly} within $4$.  (Basically, on this view, $b_\Sigma$ is treated
as an honorary member of $B_4$.)  And similarly, the permissive view
would say that
\begin{equation}
  b_1(B_3, B_{\bar{3}}, b_\sigma) \ne b_1(B_3, B_{\bar{3}},b'_\sigma)
  \label{eq-perm2}
\end{equation}
should not count as a violation of local causality since $b_\sigma$ is
at least \emph{partly} within $3$.  (Basically, on this view, $b_\sigma$ is treated
as an honorary member of $B_3$.)

On the other hand, there is also the alternative \emph{strict} view which would
answer ``no'' to  the above question.  That is, it would treat
Equation \eqref{eq-perm1} (variation of $b_\Sigma$ in response to
variations confined to $2$) as a violation of local causality since
$b_\Sigma$ is \emph{not} completely confined within region $4$, i.e.,
within the
future light cone of $2$.  (So this view treats $b_\Sigma$ as an
honorary member of $B_{\bar{4}}$.)  And
similarly, the strict view would treat Equation \eqref{eq-perm2}
(variation of events in $1$ in response to variation in $b_\sigma$)
as a violation of local causality since $b_\sigma$ is \emph{not}
complety confined within region $3$, the past light cone of $1$.  (So
this view treats $b_\sigma$ as an honorary member of $B_{\bar{3}}$.)  

(One could of course, in principle, also consider hybrid views that
are permissive about effects in the future but strict about causes in
the past, or vice versa.  But these seem less attractive.)

I'm not prepared to take a position on which view (if either) is
correct.  My primary goal is merely to raise, rather than attempt
to definitively answer, the questions posed above.  But I will report
that at present it seems most natural to demand the strict version of
dynamical locality according to which effects must be fully confined
to (rather than merely having a ``toe in'') the future light cones of
their causes, and according to which causes must be fully confined to
(rather than merely having a ``toe in'') the past light cones of their
effects.

This strict view may even be required for consistency with what we
would regard as familiar, clear-cut cases of interaction between two
local beables, if that interaction is mediated by non-local beables:
if a localized cause $a$ can influence a snake-shaped non-local beable
$b$ that merely has a toe in the future light cone of $a$, and then
$b$ can in turn influence a further localized event $c$ (which is in $b$'s
future light cone, but with $b$ merely having a toe in the past light
cone of $c$), then $a$ can (indirectly) causally influence $c$ despite
$a$ and $c$ being unambiguously space-like separated.  This seems
like the kind of thing that we should not want our concept of
dynamical locality to allow.

One last interesting point here is that, on the strict view, the
``forward'' ($F$) and ``reverse'' ($R$) versions of locality come
apart.  For example, suppose some event $b_2$ is the (for simplicity,
one and only) causal influence on
the snake-shaped $b_\Sigma$ in Figure \ref{fig-newsnakes}.  In that
case, the strict version of $F$ is violated:  the effect ($b_\Sigma$) is not fully
confined within the future light cone of the cause ($b_2$).  But
(even) the
strict version of $R$ is respected:  the cause ($b_2$) is fully within
the past light cone of the effect ($b_\Sigma$).  And similarly if
$b_\sigma$ causes some effect $b_1$ in region $1$, strict $F$ is
respected but strict $R$ is violated.  

To summarize, while there are several equivalent ways to formulate
dynamical locality for a (deterministic) theory of exclusively local
beables, ambiguities and complications arise for theories with
non-local beables.  In particular, a non-local beable may be neither
unambiguously time-like-, nor unambiguously space-like-, separated from
a local beable with which it enters into a cause-effect relation.  So the
intuitive idea depicted in Figure \ref{fig-plc} does not decide the
question of compatibility with dynamical locality.  A permissive view
(according to which causes/effects need merely have a ``toe in'' the
past/future light cones of their effects/causes) seems implausible and
potentially problematic, but we would need to be very careful if we
adopt a strict view (according to which causes/effects must be fully
confined within the past/future light cones of their effects/causes)
since, on this view, the $F$ (effects must be fully confined to the
future light cones of their causes) and $R$ (causes must be fully
confined to the past light cones of their effects) cease to be
equivalent, the way they are for TELBs.

\section{Alice, Bob, and Everett}

It has sometimes been suggested that Everettian QM is local, simply on the
grounds that it omits the collapse postulate which is the locus of
dynamical nonlocality in orthodox QM.  However, in line with Bell's dictum
(``you must identify in your theory `local \emph{be}ables'''), it was
always challenging to assess -- or even make sense of -- this claim.
How can a theory with no clear posited ontology in ordinary physical
space be said to respect relativity's prohibition on faster-than-light
influences among objects in ordinary physical space?

Wallace and Timpson's ``Spacetime State Realism'' (SSR) was thus a very
welcome contribution.  \cite{wt}  By introducing ontological elements -- beables
-- in ordinary physical space, SSR provides the pre-requisite
foundation, demanded by Bell, for meaningful discussion of EQM's
compatibility with dynamical locality.

In the context of a quantum field theory, the ontology posited by SSR
consists of -- or more precisely is described by -- the reduced
density operators (or equivalently reduced density matrices, RDMs)
associated with spacetime regions.  In the limit, we might think of
this as similar to an ordinary classical field theory, with RDMs at
each point constituting a field of local beables.  But here the value
space of the field -- a function of two real field values, or
equivalently an infinite-dimensional matrix, for a single real scalar
QFT -- is quite complicated compared to the classical case which just
assigns a single real number to each spacetime point.  This complexity
of the (mathematical description of the) local beables posited by SSR
raises questions about how/whether the theory can be
understood as empirically adequate:  whereas it is easy to identify,
e.g., a table as a table-shaped region of different-than-background field
value (in a classical field theory) or a table-shaped region of
higher-than-background particle density (in classical or Bohmian
mechanics where the local beables are simply particles with positions)
it is more challenging to find tables in a theory which assigns, to
each spacetime point, an infinite-dimensional matrix.

But let us here set this sort of worry completely aside, accept for
the sake of discussion that RDMs associated with space-time points (or
vanishingly small space-time regions) provide a viable slate of local
beables, and focus on another issue.  Namely:  these local beables do
not exhaust the spacetime ontology posited by the theory.  In addition
to the RDMs for point-like regions, SSR also posits -- as non-local
beables -- RDMs associated with larger, extended regions.

Wallace and Timpson acknowledge this when they describe SSR as
involving ``kinematical non-locality'' or ``non-separability''.  The
non-separability consists in the fact that,
\begin{quote}
``given two spacelike separated regions $A$, $B$, the state of
$A \cup B$ is given by a density operator [$\hat{\rho}_{AB}$], and
that density operator determines (via the partial trace operation) the
states $\hat{\rho}_A$ and $\hat{\rho}_B$ of A and B separately.  But
the converse is not true:  as in general in quantum mechanics,
many density operators for $A \cup B$ have the same restrictions to $A$
and to $B$.'' \cite{wt2}
\end{quote}

Earlier we mentioned the example of the total energy stored in the
electromagnetic field in some extended region and argued that this
property of the extended region should not be granted ``beable status''
because it is entirely reducible to the properties of that region's
parts, i.e., it is entirely reducible to (already-recognized) local beables.
The situation in SSR EQM is different:  there are
physically real facts about the joint region $A \cup B$ which are captured by
$\rho_{AB}$ but not by $\rho_A$ and $\rho_B$ together.  Not all properties
of the whole are determined by the properties of the parts, so the RDM
describing the state of the whole must be included in the ontology,
in addition to the RDMs describing the states of the
parts.\footnote{Actually this raises the question of ``redundancy''
  that has been pointed out previously.  (See, e.g., \cite{ground} and
  \cite{solipsism}.)  The case for
  requiring $\rho_{AB}$ as part of the ontology is that it captures
  physically real facts that are not captured by $\rho_A$ and $\rho_B$
  individually.  So if $\rho_A$ and $\rho_B$ are already posited as
  local beables, then $\rho_{AB}$ must be posited as a non-local
  beable.  But since $\rho_A$ and $\rho_B$ can be calculated from
  $\rho_{AB}$ why not deny beable status to the now-redundant
  $\rho_A$ and $\rho_B$ and
  take $\rho_{AB}$ as the only beable pertaining to regions A and B?
  The problem, of course, is that this move triggers a cascade which
  leaves one back where one started, prior to SSR, with only the universal quantum
  state $|\Psi\rangle$ or the equivalent universal density operator $\rho =
  |\Psi\rangle \langle \Psi |$ and no space-time ontology in terms of
  which one might meaningfully discuss dynamical locality, empirical adequacy, etc.}
$\rho_{AB}$, that is, must have beable status.  But it is a non-local beable,
irreducibly associated with an extended region.  

To summarize, my understanding is that SSR EQM posits, as local
beables, RDMs for pointlike regions -- \emph{and}, in addition, as non-local
beables, RDMs for larger regions (consisting of multiple space-like
separated points).  I believe this is what the authors intend as well:
Wallace writes, for example, that
\begin{quote}
``Timpson’s and my proposal ... is very straightforward:  just take
the density operator of each subsystem to represent the intrinsic
properties which that subsystem instantiates, just as the field values
assigned to each spacetime point in electromagnetism represented the
(electromagnetic) intrinsic properties which that point instantiated.” \cite{wallacebook}
\end{quote}
I see no reason why the subsystems to which the theory assigns states
should not include (e.g.) pairs of points in addition to single
points.  And the fact that the states assigned to such point-pair
subsystems is not reducible to -- is not redundant given -- the states
of the two point subsystems individually (in other words the
fact that there is non-separability) seems to imply that the states
assigned to all of the
subsystems -- both the localized point-like ones, and the
non-localized pair-of-points or snake-shaped ones -- should have the
same ``beable status''.  It's just that some of these will be local
beables and some will be nonlocal beables. 

Our goal, then, in light of the discussion in the previous section, is to
explore the implications of these non-local beables for the dynamical
locality of the theory.  

We can most easily illustrate the crucial point, without
losing anything important, in terms
of a standard EPR set up, described not in terms of QFT but rather
non-relativistic QM:  spatially-separated agents Alice and Bob
share a pair of spin-$\sfrac{1}{2}$ particles, $a$ and $b$, in the usual total-spin-zero
``singlet'' state
\begin{equation}
|\psi_0\rangle = \frac{1}{\sqrt{2}} \big( | \uparrow \rangle_a |
  \downarrow \rangle_b - |\downarrow\rangle_a |\uparrow\rangle_b \big)
\end{equation}
where $|\uparrow\rangle$ and $|\downarrow\rangle$ are, say, the one-particle
spin-up- and spin-down-along-z states.  It will be sufficient to
consider the case in which Alice and Bob are prepared to perhaps
measure the spins of their respective particles along the z-direction
so that, say, on an earlier time-slice $\sigma$ the quantum state is
\begin{eqnarray}
| \Psi_\sigma \rangle &=& |\psi_0\rangle |\text{r}\rangle_A
                          |\text{r}\rangle_B  \\
  &=& \frac{1}{\sqrt{2}} \Big( | \uparrow \rangle_a |
  \downarrow \rangle_b - |\downarrow\rangle_a |\uparrow\rangle_b
      \Big) |\text{r}\rangle_A
      |\text{r}\rangle_B \nonumber
      \label{eq-earlier}
\end{eqnarray}
where ``r'' stands for ``ready''.
And then, let's say, the quantum state on a later time-slice $\Sigma$, 
after Alice has performed her measurement and
observed its result, is
\begin{equation}
|\Psi_\Sigma \rangle = \frac{1}{\sqrt{2}} \Big( | \uparrow \rangle_a |
  \downarrow \rangle_b |\text{u}\rangle_A - |\downarrow\rangle_a
  |\uparrow\rangle_b |\text{d}\rangle_A \Big)
|\text{r}\rangle_B.
\label{eq-later}
\end{equation}
where ``u'' and ``d'' respectively denote states of Alice and her measuring
equipment in which the measurement has gone to completion and the
result has been observed to be ``up'' or ``down''. 

So that's how the quantum state (of the here-relevant degrees of
freedom) evolves.  
But what is going on in spacetime during this process according to SSR?

At the earlier time, the states of the particles $a$ and $b$
individually are described by RDMs $\rho^\sigma_a$ and
$\rho^\sigma_b$.  The partial tracing over the Alice and Bob degrees
of freedom is trivial and we are left with
\begin{eqnarray}
\rho^\sigma_a &=& \text{Tr}_b \Big( |\psi_0 \rangle \langle \psi_0 | \Big)
           \nonumber \\
  &=& \langle \uparrow |_b \Big( |\psi_0 \rangle \langle \psi_0 |
      \Big) | \uparrow \rangle_b +\langle \downarrow |_b \Big( |\psi_0 \rangle \langle \psi_0 |
      \Big) | \downarrow \rangle_b \nonumber \\
  &=& \frac{1}{2} \Big( |\uparrow\rangle \langle \uparrow | +
      |\downarrow \rangle \langle \downarrow | \Big) \nonumber \\
  &\rightarrow& \frac{1}{2} \left( \begin{array}{cc} 1 & 0 \\ 0 &
                                                           1 \end{array} \right)
\end{eqnarray}
where, in the last line, we have represented the operator as a matrix in the
$|\uparrow\rangle$, $|\downarrow\rangle$ basis.  
The state of Bob's particle is similarly
\begin{equation}
\rho^\sigma_b \rightarrow \frac{1}{2} \left( \begin{array}{cc} 1 & 0 \\ 0 &
                                                           1 \end{array} \right).
\end{equation}
And the state of the (spatially extended) joint two-particle system is
given by
\begin{eqnarray}
  \rho^\sigma_{ab} &=& |\psi_0 \rangle \langle \psi_0 | \nonumber \\
  &\rightarrow& \frac{1}{2} \left( \begin{array}{cccc} 0&0&0&0\\ 0&1&-1&0 \\
                               0&-1&1&0 \\ 0&0&0&0 \end{array} \right)
\end{eqnarray}
where we have here represented the operator in the $|\uparrow\rangle_a
|\uparrow\rangle_b$, $|\uparrow\rangle_a |\downarrow\rangle_b$,
$|\downarrow\rangle_a |\uparrow\rangle_b$, $|\downarrow\rangle_a
|\downarrow\rangle_b$ basis.  
The presence of the off-diagonal elements here indicates, in the usual
way, that the two particle state is not merely a mixture of $|\uparrow
\downarrow\rangle$ and $|\downarrow\uparrow\rangle$, but is a coherent
(entangled) superposition.

What about the situation later, on $\Sigma$, after Alice has performed
her measurement and the quantum state is given by Equation
\eqref{eq-later}?  A quick calculation reveals that the RDMs for the
individual particles are unchanged:
\begin{equation}
\rho^\Sigma_a = \rho^\sigma_a \rightarrow \frac{1}{2}
\left( \begin{array}{cc} 1&0 \\ 0 & 1 \end{array} \right)
\end{equation}
and
\begin{equation}
\rho^\Sigma_b = \rho^\sigma_b \rightarrow \frac{1}{2}
\left( \begin{array}{cc} 1&0 \\ 0 & 1 \end{array} \right).
\end{equation}
But the (non-local beable) RDM describing the joint two-particle
system is different on $\Sigma$ than it was on $\sigma$:
\begin{equation}
  \rho^\Sigma_{ab} \rightarrow
  \frac{1}{2} \left( \begin{array}{cccc} 0&0&0&0\\ 0&1&0&0 \\
                               0&0&1&0 \\ 0&0&0&0 \end{array} \right).
\end{equation}
The difference here is just an elementary case of decoherence:  the
interaction (and resulting entanglement) between the two-particle
system and Alice's measuring apparatus has killed the off-diagonal
entries.

The now-standard case for the SSR version of Everettian QM
being a dynamically local theory is illustrated
here by the fact that Alice's intervention to measure the z-spin of
her particle ($a$) does not affect the local beable  associated with
Bob's distant particle ($b$):  $\rho^\sigma_b = \rho^\Sigma_b$.   Or,
expressing this in terms of Vaidman's formulations, we may consider
two evolutions:  the one already described (in which Alice performs a
measurement of the z-spin of her particle), and another in which she
decides not to perform that measurement so that, say,
$|\Psi'_\Sigma\rangle = |\Psi_\sigma\rangle$.  Suppose Alice's decision (to
measure, or not) occurs in region $2$ of Figure \ref{fig-lc}, with
everything in $\bar{2}$ being the same; then dynamical locality
requires the two solutions to only differ, on $\Sigma$, in region
$4$.  So in particular they must be identical in $1$.  And they are:
the local beable $\rho^\Sigma_b$ capturing the state of Bob's particle
is the same whether Alice measures her particle or not.  Distant local
beables are not affected (right away) by localized nearby
interventions.

But of course the interesting point is that the non-local beable
describing the joint state $\rho_{ab}$ of the two-particle system \emph{is}
affected (right away) by Alice's measurement intervention.  And while
part of this two-particle system is of course local to Alice, the
other part is not.  So we seem to have a situation exactly like those
discussed in the previous section.  Indeed, the non-local beable
(described by) $\rho_{ab}$ seems to behave precisely like the
mysterious holistic property of the snake that we considered in the
introduction and like ``Case (i)'' from just after Figure
\ref{fig-newsnakes} in Section \ref{sec2}.

The fact that the RDM of a spatially-extended region can be affected
by an intervention localized to some part of the region has been
pointed out before.  But it has almost always been dismissed as not
conflicting with dynamical locality.  Most often this dismissal has
been based on the idea that insofar as $\rho_{ab}$ pertains to the
localized region of Bob's particle, it describes merely relational or
extrinsic properties of that particle, such that the change in
$\rho_{ab}$ does not imply any physical change outside the future
light cone of Alice's measurement.  This argument will be discussed
further in the following section. 

Before turning to that, however, let us briefly engage the recent commentary
of Chua and Sebens on this sort of situation.  They begin by summarizing
essentially the same example we've given here:
\begin{quote}
 ``One might worry that non-separability could lead to violations of
  relativistic causality, because the non-separable states of widely
  spread-out composite systems can immediately change when acted upon
  at one location.  ....  A local action within $A$ has changed the
  global state of the two particles, a state that extends into a
  region, $B$, that is space-like separated from $A$.  Furthermore, note
  that the reduced density matrices of the two particles are the same
  [before and after Alice's measurement].  This local action leaves
  the local states of the $A$-particle and the $B$-particle unchanged
  while causing global changes in the state of the two particles
  together.''
  \end{quote}
  They then continue:
  \begin{quote}
  ``Although this might appear to be a violation of relativistic
  locality, it is not.  The global state of the two particles, located
  in region
  $A \cup B$, has changed due to a cause within region $A$, a part of
  $A \cup B$.
  The cause is local to the effect.'' \cite{cs}
\end{quote}
Visualizing this in terms of our Figure \ref{fig-newsnakes},
they mean, I believe, that Alice's intervention (localized, say, in
$2$) is fully confined to the past light cone of the affected
(snake-shaped) entity $b_\Sigma$.

That, of course, is true.  In this kind of case, the strict reading of
the ``reverse'' version ($R$) of dynamical locality is respected.

But as we discussed at the end of the last section, the ``forward''
($F$) and ``reverse'' ($R$) versions of dynamical locality come apart
when we are no longer dealing with a TELB.  And the example here is
precisely the sort discussed previously, in which (strict) $R$ is
respected but (strict) $F$ is violated.  So
while it is true that, in this kind of situation, the cause is fully
confined to the past light cone of the effect, it does not follow that
-- and it is simply not true that -- the effect is fully confined to
the future light cone of the cause.  

``The cause'' may be ``local to the effect'', as Chua and Sebens claim,
but the effect is not local to the cause!  

The difficulty can also be seen in a relatively direct way if we
return to QFT.  Wallace and Timpson argue for the very general claim that ``quantum field
theory is dynamically local on the Everett interpretation'' \cite{wt2}
on the following grounds.  If density operators $\hat{\rho}$ and
$\hat{\rho}'$ (characterizing the state of things on a hyperplane like
$\sigma$ in our Figure \ref{fig-lc}) disagree within region $2$ but
agree everywhere else, then $\hat{\rho}' = \hat{U} \hat{\rho} \,
\hat{U}^\dagger$ for some $\hat{U}$ constructed exclusively from
operators localized in region $2$.  But then if $\hat{X}$ represents a
property  in region $1$ (spacelike separated from $2$),
$\text{Tr} \left( \hat{\rho}' \hat{X} \right) = \text{Tr} \left(
  \hat{U} \hat{\rho} \, \hat{U}^\dagger \hat{X} \right) = \text{Tr}
\left( \hat{\rho} \hat{X} \right)$ since, in a relativistic QFT we
would have $\left[ \hat{U},\hat{X} \right] =0$ due to the space-like
separation.  Therefore the value of $X$ in region $1$ is independent
of the intervention in $2$.

However, if we replace $\hat{X}$ with some non-localized $\hat{Y}$,
built, say, from operators living within the snake-shaped region occupied by
$b_\Sigma$ in Figure \ref{fig-newsnakes}, then there is no reason for
$\left[ \hat{U}, \hat{Y} \right]$ to vanish, and hence no reason why
the value of $Y$ should be independent of the intervention in $2$.
Localized interventions \emph{can}, in ``quantum field theory ... on
the Everett interpretation'', affect ``holistic'' properties (like
that misleadingly depicted as the wavy vs spiky stripe pattern in
Figure \ref{fig-snake2}) which are not fully confined to the future
light cone of the intervention.

So I think this kind of situation poses a much more serious problem
than has previously been recognized for the developing consensus view
of SSR EQM as a non-separable but dynamically local theory.

\section{Discussion}

As mentioned in the last section, it has become common to regard the
fact that, for example, Alice can affect $\rho_{ab}$ (which describes
the state of the joint system consisting of two spatially separated
particles) with an intervention localized around just particle $a$,
as perfectly compatible with dynamical locality, basically on the
grounds that $\rho_{ab}$ merely describes relational properties of $a$
and $b$.

Alyssa Ney develops an analogy to
Socrates and his wife, Xanthippe.  \cite{ney}  As soon as Socrates
drinks the hemlock and tragically dies, Xanthippe immediately becomes
a widow despite being, let's assume, far away.  But surely there is no
conflict with dynamical locality in this kind of case:  Xanthippe's
marital status is not an intrinsic property she possesses -- not
something reflected in the local beables composing her person -- but is
instead a purely relational property, grounded in and fully reducible
to the intrinsic properties of Socrates and Xanthippe individually.
Indeed, in this case, the
relational property changes exclusively as a result of the changes in
the local beables composing Socrates.

In the same way, it is suggested, when Alice measures the spin of
particle $a$, the intrinsic properties of Bob's distant particle $b$,
as described by $\rho_b$, do not change.  The joint two-particle state
$\rho_{ab}$ does change, but this, it is suggested, just describes
relationships between the two sub-systems.  And, like Xanthippe's
marital status, there is, supposedly, no reason to insist that
dynamical locality prevents such relational properties from being affected by interventions
that act directly on just one of the sub-systems.

This same basic argument has been around for some time.  
Timpson and Brown, for example, considered this kind of case already
in a 2002 paper.  They claim
that the ``appearance of non-locality is ... not genuine'' on the
following grounds:
\begin{quote}
``What have changed as a result of Alice’s measurement are the
relative states of Bob’s system; that is, roughly, relational
properties of his system.  It is no mystery that relational properties
can be affected unilaterally by operations on one of the \emph{relata} and it
certainly does not connote non-locality.  (Compare `$x$ is heavier
than $y$'; we might make $y$ heavier by adding weights, so that this
statement becomes false, but this would not indicate a non-local
effect on $x$.)'' \cite{tb}
\end{quote}
They go on to state that ``the genuine change is in fact all on
Alice’s side.''

This last, at least in the example we considered here, is surely false:  there are, of course, changes
on Alice's side if/when Alice performs her measurement.  But a change
to the local beable description of her particle is simply not among them:
$\rho^\Sigma_a = \rho^\sigma_a$.

Alice's measurement affects the
global state $\rho_{ab}$ of the two-particle system despite failing to
affect the states $\rho_a$ and $\rho_b$ of \emph{either} of the two particles
separately.  The fact that Alice's intervention doesn't affect
$\rho_b$ is touted as the main reason for claiming the theory is
dynamically local.  But the fact that $\rho_a$ is also unaffected
significantly undermines the case for thinking of $\rho_{ab}$ as merely
describing relationships between the $a$ and $b$ systems.  In
the Socrates/Xanthippe example, the relational property changes \emph{only}
due to a change in the intrinsic properties of one of the related
sub-systems:  when Socrates
dies, the only genuine physical change is to Socrates.  The
relationship between Socrates and Xanthippe changes \emph{exclusively}
because of this change in Socrates.  The relationship in this example is
entirely reducible to the intrinsic properties of the relata  -- it is
not an additional, non-local beable -- and this
is why there is no conflict with dynamical locality.

But the situation with Alice and the particles, according to SSR EQM,
is, as we have argued, completely different.
The fact that $\rho_{ab}$ changes without any concomitant changes in
$\rho_a$ or $\rho_b$ means that thinking of $\rho_{ab}$ as merely
describing ``relational properties'', in anything like the sense
suggested by the Socrates/Xanthippe analogy, is at best highly
misleading.  However we want to think about $\rho_{ab}$, the crucial
point is that it is not reducible to the properties of $a$ and $b$ --
that, indeed, is why we felt obligated to grant ``beable status'' to
$\rho_{ab}$, i.e., to supplement the local beables $\rho_a$ and
$\rho_b$ with this additional mysterious non-local beable.  

But if it's right that things like $\rho_{ab}$ have beable status, we need to take
them seriously as aspects of the ontology (even if they are a bit cryptic and
unfamiliar) and not let ourselves be tempted into inappropriately
pigeonholing them into some more familiar category such as
``(reducible) relational property''.  

In theories that posit them, non-local beables are
physical entities, every bit as \emph{real} as
local beables, but irreducibly occupying extended regions.  Wallace
suggests that ``picturesquely, we can think of [a pair of
spatially-separated but entangled particles as having] a string
connecting those states, representing the nonlocal relation between
them.''  \cite{wallacebook}  That, I think, is a perfectly good
metaphor.  And it means that such a pair (including the metaphorical
string connecting them) occupies a snake-shaped region which
irreducibly bears global properties that can be affected by localized
interventions.  In the concrete case we have focused on, this means that
the effects of a localized intervention need not be confined to the
future light cone of that intervention.

We have a violation of the
``strict'' reading of Vaidman's ``forward'' ($F$) version of dynamical
locality.

I mentioned in the introduction that a number of people (including at least
Rodi Tumulka, Cai Waegell, Kelvin McQueen, and myself) have suggested
that the apparent correlatedness of distant Everettian world-splittings seemed
to possess an air of dynamical nonlocality, even though the local
beables on both sides, as characterized in SSR, behaved
independently.  For me at least, the hard-to-articulate appearance of an ``air of
dynamical nonlocality'' arose precisely because the two sets of local
beables, together, \emph{miss} -- \emph{leave out} -- the
connectedness, apparent in the structure of the universal quantum
state, of the two spatially-separated world-splittings.

But this connectedness is also captured in the RDMs for spatially-extended
systems which seemingly possess (non-local) beable status
in the SSR theory.  Bringing those nonlocal beables -- and
\emph{their} cause-effect relations -- explicitly into the discussion
seems to raise interesting and difficult new questions for those
defending EQM as a dynamically local theory and in particular seems to bring
out the ``air of dynamical nonlocality'' (previously gestured at in
terms of the correlatedness of world-splittings) a bit
more crisply.

\vspace{.2in}

{\bf{Acknowledgements:}}  Thanks to Alyssa Ney, Dustin Lazarovici,
Ward Struyve, Cai Waegell, and Ken Wharton for stimulating
conversations and/or helpful comments on an
early draft.

\end{document}

%% file: figures/snake.pdf_tex
\begingroup%
  \makeatletter%
  \providecommand\color[2][]{%
    \errmessage{(Inkscape) Color is used for the text in Inkscape, but the package 'color.sty' is not loaded}%
    \renewcommand\color[2][]{}%
  }%
  \providecommand\transparent[1]{%
    \errmessage{(Inkscape) Transparency is used (non-zero) for the text in Inkscape, but the package 'transparent.sty' is not loaded}%
    \renewcommand\transparent[1]{}%
  }%
  \providecommand\rotatebox[2]{#2}%
  \newcommand*\fsize{\dimexpr\f@size pt\relax}%
  \newcommand*\lineheight[1]{\fontsize{\fsize}{#1\fsize}\selectfont}%
  \ifx\svgwidth\undefined%
    \setlength{\unitlength}{1585.2849884bp}%
    \ifx\svgscale\undefined%
      \relax%
    \else%
      \setlength{\unitlength}{\unitlength * \real{\svgscale}}%
    \fi%
  \else%
    \setlength{\unitlength}{\svgwidth}%
  \fi%
  \global\let\svgwidth\undefined%
  \global\let\svgscale\undefined%
  \makeatother%
  \begin{picture}(1,0.50053285)%
    \lineheight{1}%
    \setlength\tabcolsep{0pt}%
    \put(0,0){\includegraphics[width=\unitlength,page=1]{snake.pdf}}%
  \end{picture}%
\endgroup%

%% file: figures/snake2.pdf_tex
\begingroup%
  \makeatletter%
  \providecommand\color[2][]{%
    \errmessage{(Inkscape) Color is used for the text in Inkscape, but the package 'color.sty' is not loaded}%
    \renewcommand\color[2][]{}%
  }%
  \providecommand\transparent[1]{%
    \errmessage{(Inkscape) Transparency is used (non-zero) for the text in Inkscape, but the package 'transparent.sty' is not loaded}%
    \renewcommand\transparent[1]{}%
  }%
  \providecommand\rotatebox[2]{#2}%
  \newcommand*\fsize{\dimexpr\f@size pt\relax}%
  \newcommand*\lineheight[1]{\fontsize{\fsize}{#1\fsize}\selectfont}%
  \ifx\svgwidth\undefined%
    \setlength{\unitlength}{1585.2849884bp}%
    \ifx\svgscale\undefined%
      \relax%
    \else%
      \setlength{\unitlength}{\unitlength * \real{\svgscale}}%
    \fi%
  \else%
    \setlength{\unitlength}{\svgwidth}%
  \fi%
  \global\let\svgwidth\undefined%
  \global\let\svgscale\undefined%
  \makeatother%
  \begin{picture}(1,0.50053285)%
    \lineheight{1}%
    \setlength\tabcolsep{0pt}%
    \put(0,0){\includegraphics[width=\unitlength,page=1]{snake2.pdf}}%
  \end{picture}%
\endgroup%

%% file: figures/plc.pdf_tex
\begingroup%
  \makeatletter%
  \providecommand\color[2][]{%
    \errmessage{(Inkscape) Color is used for the text in Inkscape, but the package 'color.sty' is not loaded}%
    \renewcommand\color[2][]{}%
  }%
  \providecommand\transparent[1]{%
    \errmessage{(Inkscape) Transparency is used (non-zero) for the text in Inkscape, but the package 'transparent.sty' is not loaded}%
    \renewcommand\transparent[1]{}%
  }%
  \providecommand\rotatebox[2]{#2}%
  \newcommand*\fsize{\dimexpr\f@size pt\relax}%
  \newcommand*\lineheight[1]{\fontsize{\fsize}{#1\fsize}\selectfont}%
  \ifx\svgwidth\undefined%
    \setlength{\unitlength}{1051.95959473bp}%
    \ifx\svgscale\undefined%
      \relax%
    \else%
      \setlength{\unitlength}{\unitlength * \real{\svgscale}}%
    \fi%
  \else%
    \setlength{\unitlength}{\svgwidth}%
  \fi%
  \global\let\svgwidth\undefined%
  \global\let\svgscale\undefined%
  \makeatother%
  \begin{picture}(1,0.53286459)%
    \lineheight{1}%
    \setlength\tabcolsep{0pt}%
    \put(0.3189646,0.26033995){\color[rgb]{0,0,0}\transparent{0.99607801}\makebox(0,0)[lt]{\lineheight{1.25}\smash{\begin{tabular}[t]{l}$\text{1}$\end{tabular}}}}%
    \put(0.8002655,0.26260541){\color[rgb]{0,0,0}\transparent{0.99607801}\makebox(0,0)[lt]{\lineheight{1.25}\smash{\begin{tabular}[t]{l}$\text{2}$\end{tabular}}}}%
    \put(0.26719887,0.15135221){\color[rgb]{0,0,0}\transparent{0.99607801}\makebox(0,0)[lt]{\lineheight{1.25}\smash{\begin{tabular}[t]{l}$\text{causes}$\end{tabular}}}}%
    \put(0.76853852,0.14420338){\color[rgb]{0,0,0}\transparent{0.99607801}\makebox(0,0)[lt]{\lineheight{1.25}\smash{\begin{tabular}[t]{l}$\sm \text{space}$\end{tabular}}}}%
    \put(0.77711834,0.3654744){\color[rgb]{0,0,0}\transparent{0.99607801}\makebox(0,0)[lt]{\lineheight{1.25}\smash{\begin{tabular}[t]{l}$\sm \text{time}$\end{tabular}}}}%
    \put(0.25955721,0.37031464){\color[rgb]{0,0,0}\transparent{0.99607801}\makebox(0,0)[lt]{\lineheight{1.25}\smash{\begin{tabular}[t]{l}$\text{effects}$\end{tabular}}}}%
    \put(0,0){\includegraphics[width=\unitlength,page=1]{plc.pdf}}%
  \end{picture}%
\endgroup%

%% file: figures/lc.pdf_tex
\begingroup%
  \makeatletter%
  \providecommand\color[2][]{%
    \errmessage{(Inkscape) Color is used for the text in Inkscape, but the package 'color.sty' is not loaded}%
    \renewcommand\color[2][]{}%
  }%
  \providecommand\transparent[1]{%
    \errmessage{(Inkscape) Transparency is used (non-zero) for the text in Inkscape, but the package 'transparent.sty' is not loaded}%
    \renewcommand\transparent[1]{}%
  }%
  \providecommand\rotatebox[2]{#2}%
  \newcommand*\fsize{\dimexpr\f@size pt\relax}%
  \newcommand*\lineheight[1]{\fontsize{\fsize}{#1\fsize}\selectfont}%
  \ifx\svgwidth\undefined%
    \setlength{\unitlength}{1166.0450592bp}%
    \ifx\svgscale\undefined%
      \relax%
    \else%
      \setlength{\unitlength}{\unitlength * \real{\svgscale}}%
    \fi%
  \else%
    \setlength{\unitlength}{\svgwidth}%
  \fi%
  \global\let\svgwidth\undefined%
  \global\let\svgscale\undefined%
  \makeatother%
  \begin{picture}(1,0.4807293)%
    \lineheight{1}%
    \setlength\tabcolsep{0pt}%
    \put(0,0){\includegraphics[width=\unitlength,page=1]{lc.pdf}}%
    \put(0.34133164,0.10900665){\color[rgb]{0,0,0}\transparent{0.99607801}\makebox(0,0)[lt]{\lineheight{1.25}\smash{\begin{tabular}[t]{l}$\sm 3$\end{tabular}}}}%
    \put(0.72592289,0.12408631){\color[rgb]{0,0,0}\transparent{0.99607801}\makebox(0,0)[lt]{\lineheight{1.25}\smash{\begin{tabular}[t]{l}$\sm 2$\end{tabular}}}}%
    \put(0,0){\includegraphics[width=\unitlength,page=2]{lc.pdf}}%
    \put(0.34137246,0.35647764){\color[rgb]{0,0,0}\transparent{0.99607801}\makebox(0,0)[lt]{\lineheight{1.25}\smash{\begin{tabular}[t]{l}$\sm 1$\end{tabular}}}}%
    \put(0.72146288,0.34636305){\color[rgb]{0,0,0}\transparent{0.99607801}\makebox(0,0)[lt]{\lineheight{1.25}\smash{\begin{tabular}[t]{l}$\sm 4$\end{tabular}}}}%
    \put(0.05418556,0.35216311){\color[rgb]{0,0,0}\transparent{0.99607801}\makebox(0,0)[lt]{\lineheight{1.25}\smash{\begin{tabular}[t]{l}$\Sigma$\end{tabular}}}}%
    \put(0.06218642,0.11609575){\color[rgb]{0,0,0}\transparent{0.99607801}\makebox(0,0)[lt]{\lineheight{1.25}\smash{\begin{tabular}[t]{l}$\sigma$\end{tabular}}}}%
  \end{picture}%
\endgroup%

%% file: figures/newsnakes.pdf_tex
\begingroup%
  \makeatletter%
  \providecommand\color[2][]{%
    \errmessage{(Inkscape) Color is used for the text in Inkscape, but the package 'color.sty' is not loaded}%
    \renewcommand\color[2][]{}%
  }%
  \providecommand\transparent[1]{%
    \errmessage{(Inkscape) Transparency is used (non-zero) for the text in Inkscape, but the package 'transparent.sty' is not loaded}%
    \renewcommand\transparent[1]{}%
  }%
  \providecommand\rotatebox[2]{#2}%
  \newcommand*\fsize{\dimexpr\f@size pt\relax}%
  \newcommand*\lineheight[1]{\fontsize{\fsize}{#1\fsize}\selectfont}%
  \ifx\svgwidth\undefined%
    \setlength{\unitlength}{1166.0450592bp}%
    \ifx\svgscale\undefined%
      \relax%
    \else%
      \setlength{\unitlength}{\unitlength * \real{\svgscale}}%
    \fi%
  \else%
    \setlength{\unitlength}{\svgwidth}%
  \fi%
  \global\let\svgwidth\undefined%
  \global\let\svgscale\undefined%
  \makeatother%
  \begin{picture}(1,0.4807293)%
    \lineheight{1}%
    \setlength\tabcolsep{0pt}%
    \put(0,0){\includegraphics[width=\unitlength,page=1]{newsnakes.pdf}}%
    \put(0.34133164,0.10900665){\color[rgb]{0,0,0}\transparent{0.99607801}\makebox(0,0)[lt]{\lineheight{1.25}\smash{\begin{tabular}[t]{l}$\sm 3$\end{tabular}}}}%
    \put(0.72592289,0.12408631){\color[rgb]{0,0,0}\transparent{0.99607801}\makebox(0,0)[lt]{\lineheight{1.25}\smash{\begin{tabular}[t]{l}$\sm 2$\end{tabular}}}}%
    \put(0,0){\includegraphics[width=\unitlength,page=2]{newsnakes.pdf}}%
    \put(0.34137246,0.35647764){\color[rgb]{0,0,0}\transparent{0.99607801}\makebox(0,0)[lt]{\lineheight{1.25}\smash{\begin{tabular}[t]{l}$\sm 1$\end{tabular}}}}%
    \put(0.72146288,0.34636305){\color[rgb]{0,0,0}\transparent{0.99607801}\makebox(0,0)[lt]{\lineheight{1.25}\smash{\begin{tabular}[t]{l}$\sm 4$\end{tabular}}}}%
    \put(0.05418556,0.35216311){\color[rgb]{0,0,0}\transparent{0.99607801}\makebox(0,0)[lt]{\lineheight{1.25}\smash{\begin{tabular}[t]{l}$\Sigma$\end{tabular}}}}%
    \put(0.06218642,0.11609575){\color[rgb]{0,0,0}\transparent{0.99607801}\makebox(0,0)[lt]{\lineheight{1.25}\smash{\begin{tabular}[t]{l}$\sigma$\end{tabular}}}}%
    \put(0,0){\includegraphics[width=\unitlength,page=3]{newsnakes.pdf}}%
    \put(0.54727942,0.3450295){\color[rgb]{0,0,0}\transparent{0.99607801}\makebox(0,0)[lt]{\lineheight{1.25}\smash{\begin{tabular}[t]{l}$\sm b_\Sigma$\end{tabular}}}}%
    \put(0.53553954,0.09307582){\color[rgb]{0,0,0}\transparent{0.99607801}\makebox(0,0)[lt]{\lineheight{1.25}\smash{\begin{tabular}[t]{l}$\sm b_\sigma$\end{tabular}}}}%
  \end{picture}%
\endgroup%